\documentclass[iop]{emulateapj}
\usepackage{graphicx}
\usepackage[flushleft]{threeparttable}
\usepackage{blindtext}
\usepackage{amsmath}
\usepackage{mathtools}
\usepackage{multirow}
\usepackage{hyperref}
\hypersetup{
	colorlinks	= true,
	linkcolor	= red,
	urlcolor	= cyan,
	citecolor	= blue
}

\newcommand{\taurex}{\mbox{$\mathcal{T}$-REx} }
\newcommand{\taurexng}{\mbox{$\mathcal{T}$-REx}}

\begin{document}

\title{Detection of an atmosphere around the super-Earth 55 Cancri e}

\author{A. Tsiaras, M. Rocchetto, I. P. Waldmann}
\affil{Department of Physics \& Astronomy, University College London, Gower Street, WC1E6BT London, United Kingdom}
\author{O. Venot}
\affil{Instituut voor Sterrenkunde, Katholieke Universiteit Leuven, Celestijnenlaan 200D, 3001 Leuven, Belgium}
\author{R. Varley, G. Morello, M. Damiano, G. Tinetti, E. J. Barton, S. N. Yurchenko , J. Tennyson}
\affil{Department of Physics \& Astronomy, University College London, Gower Street, WC1E6BT London, United Kingdom}
\email{angelos.tsiaras.14@ucl.ac.uk}

\begin{abstract}
We report the analysis of two new spectroscopic observations of the super-Earth 55 Cancri e, in the near infrared, obtained with the WFC3 camera onboard the HST. 55 Cancri e orbits so close to its parent star, that temperatures much higher than 2000 K are expected on its surface. Given the brightness of 55 Cancri, the observations were obtained in scanning mode, adopting a very long scanning length and a very high scanning speed. We use our specialized pipeline to take into account systematics introduced by these observational parameters when coupled with the geometrical distortions of the instrument. We measure the transit depth per wavelength channel with an average relative uncertainty of 22 ppm per visit and find modulations that depart from a straight line model with a 6$\sigma$ confidence level. These results suggest that 55 Cancri e is surrounded by an atmosphere, which is probably hydrogen-rich. Our fully Bayesian spectral retrieval code, \taurexng, has identified HCN to be the most likely molecular candidate able to explain the features at 1.42 and 1.54 $\mu$m. While additional spectroscopic observations in a broader wavelength range in the infrared will be needed to confirm the HCN detection, we discuss here the implications of such result. Our chemical model, developed with combustion specialists, indicates that relatively high mixing ratios of HCN may be caused by a high C/O ratio. This result suggests this super-Earth is a carbon-rich environment even more exotic than previously thought.
\end{abstract}

\maketitle

\section{INTRODUCTION} \label{sec:introduction}

Transiting exoplanets are an invaluable source of information as they enable the measurement of a large number of orbital and planetary parameters. Especially when observed at different wavelengths, the thermal properties and composition of the exoplanetary atmosphere can be revealed \citep[e.g.][]{Seager2000, Brown2001, Tinetti2007a}.

Using the transit technique, many teams have discovered molecular features in the atmospheres of exoplanets, both from the ground \citep[e.g.][]{Redfield2008, Swain2010, Waldmann2012, Snellen2008, Snellen2010, Snellen2014, Bean2010, Bean2013} and from space \citep[e.g.][]{Charbonneau2002, Charbonneau2005,  Deming2005, Deming2011, Knutson2007, Knutson2008, Tinetti2007b, Tinetti2010, Swain2008, Swain2009a, Swain2009b,  Swain2013}.

Super-Earths are an intriguing class of planets as they do not exist within our solar system. Even basic information about their density seems to suggest that there is a variety of cases \citep[e.g.][]{Sotin2007, Grasset2009, Valencia2007, Valencia2013, Zeng2014}. According to Kepler and radial velocity statistics, super-Earths are the most abundant planets, especially around late-type stars \citep[e.g.][]{Mayor2011, Howard2012, Fressin2013, Dressing2013}. The Wide Field Camera 3 (WFC3) onboard the Hubble Space Telescope (HST), combined with the recently implemented spatial scanning technique, allows the spectroscopic observation of super-Earths, which is unprecedented. Published observations of two super-Earths, GJ1214b and HD97658b, do not show any evident transit depth modulation with wavelength \citep{Kreidberg2014a, Knutson2014b}, suggesting an atmosphere covered by thick clouds or made of molecular species much heavier than hydrogen.

At a distance of only 12 pc, 55 Cancri is an extremely interesting planetary system hosting five planets, all discovered via radial velocity measurements \citep{Butler1997, Marcy2002, McArthur2004,  Fischer2008}. Amongst them, 55 Cancri e is an ``exotic'' example of a super-Earth as it orbits very close to the host star and consequently the temperature on its surface is high, i.e. hotter than 2000 K. The initially reported values for  its period and minimum mass were 2.808 days and 14.21 $\pm$ 2.91 $M_\oplus$ respectively \citep{McArthur2004}. However \cite{Dawson2010} revised these values, reporting a period of 0.7365 days and a minimum mass of 8.3 $\pm$ 0.3 $M_\oplus$.

Transits from space of 55 Cancri e were observed with the Spitzer Space Telescope \citep{Demory2011} and MOST space telescope \citep{Winn2011}, confirming its small size ($\sim \, 2 \, R_\oplus$). The eclipse depth of 131 $\pm$ 28 ppm observed by \cite{Demory2012} with Spitzer/IRAC at 4.5 $\mu$m converts into a brightness temperature of 2360 $\pm$ 300 K. Also, repeated eclipse observations suggested a variability of the thermal emission from the dayside of the planet over time \citep{Demory2015}.  Table \ref{tab:parameters} summarizes the most recent parameters for 55 Cancri e. Given these parameters, \cite{Zeng2014} have modeled the plausible structure of the interior and \cite{Hu2014} possible atmospheric compositions. These authors together with others \citep[e.g.][]{Stevenson2013, Forget2014}, acknowledge that a gaseous envelope made of H$_2$ and He might have been retained from the protoplanetary disk. 

\begin{table}
	\small
	\center
	\caption{Observationally determined parameters of 55 Cancri e}
	\label{tab:parameters}
	\begin{tabular}{c | c }
		
		\hline \hline
		\multicolumn{2}{c}{Stellar parameters } 											\\ [0.1ex]
		\hline
		$\mathrm{[Fe/H]} \, \mathrm{[dex]}$ $^{(1)}$	& 0.31 $\pm$ 0.04						\\
		$T_\mathrm{eff} \, \mathrm{[K]}$ $^{(2)}$		& 5196 $\pm$ 24						\\
		$M_* \, [M_{\odot}]$ $^{(2)}$				& 0.905 $\pm$ 0.015						\\
		$R_* \, [R_{\odot}]$ $^{(2)}$				& 0.943 $\pm$ 0.010						\\
		$\log(g_*) \, \mathrm{[cgs]}$ $^{(2)}$ 		& 4.45 $\pm$ 0.001						\\ [1.0ex]
		
		\hline \hline
		\multicolumn{2}{c}{Planetary parameters}											\\ [0.1ex]
		\hline
		$T_\mathrm{eq} \, \mathrm{[K]}$ $^{(3)}$ 		& 1950$_{-190}^{+260}$					\\
		$M_\mathrm{p} \, [M_\oplus]$ $^{(4)}$		& 8.09 $\pm$ 0.26						\\
		$R_\mathrm{p} \, [R_\oplus]$ $^{(5)}$		& 1.990$_{-0.080}^{+0.084}$				\\
		$a \, \mathrm{[AU]}$ $^{(5)}$ 				& 0.01545$_{-0.00024}^{+0.00025}$			\\ [1.0ex]
		
		\hline\hline
		\multicolumn{2}{c}{Transit parameters}											\\ [0.1ex]
		\hline
		$T_0 \, \mathrm{[BJD]}$ $^{(5)}$			& 2455962.0697$_{-0.0018}^{+0.0017}$		\\
		$\mathrm{Period} \,  \mathrm{[days]}$ $^{(5)}$ 	& 0.7365417$_{-0.0000028}^{+0.0000025}$	\\
		$R\mathrm{p}/R_*$ $^{(5)}$ 				& 0.01936$_{-0.00075}^{+0.00079}$			\\
		$a/R_*$ $^{(5)}$						& 3.523$_{-0.040}^{+0.042}$				\\
		$i \, \mathrm{[deg]}$ $^{(5)}$ 				& 85.4$_{-2.1}^{+2.8}$					\\ [1.0ex]
		\hline
		
		\multicolumn{2}{l}{$^{(1)}$\cite{Valenti2005}, $^{(2)}$\cite{vonBraun2011}}					\\
		\multicolumn{2}{l}{$^{(3)}$\cite{Crossfield2012}, $^{(4)}$\cite{Nelson2014}}				\\
		\multicolumn{2}{l}{$^{(5)}$\cite{Dragomir2014}}										
								
	\end{tabular}
\end{table}

In this work we analyze recent observations of the transit of 55 Cancri e obtained with the WFC3 camera onboard the Hubble Space Telescope. Since the host star is very bright, these observations were obtained using spatial scans. This technique allows the telescope to slew during the exposure to avoid the saturation of the detector and has already been successfully used to provide an increasing number of exoplanetary spectra \citep[e.g.][]{Deming2013, Knutson2014a, Knutson2014b, Kreidberg2014b, Kreidberg2014a, Kreidberg2015, McCullough2014, Crouzet2014, Fraine2014, Stevenson2014}.

The scan rate used for the observations of 55 Cancri e (4.8 $''$/s) results in a very long scan of approximately 350 pixels. Typical scan lengths in currently published datasets vary between only 20 \citep[XO1b - scan rate of 0.05 $''$/s,][]{Deming2013} and 170 pixels \citep[HD97658b - scan rate of 1.4 $''$/s,][]{Knutson2014b}. This configuration can generate position-dependent systematics which need to be carefully removed. Our WFC3 scanning-mode dedicated pipeline \citep{Tsiaras2015} was created to process such challenging datasets and it is able to correct the effects of the geometric distortions. It can therefore provide a uniform analysis of different kinds of spatially scanned observations (Section \ref{sub:extraction}). Lastly, we use a range of techniques to correct for the time-dependent instrumental systematics known as the ``ramps'' (Section \ref{sub:fitting}), and calculate the wavelength-dependent transit depth.

The interpretation of the observed spectrum is carried out using our Bayesian spectral retrieval framework, \taurexng, described in \cite{Waldmann2015a,Waldmann2015b}. In particular we use the WFC3 data to constrain key atmospheric parameters such as the pressure at the surface, the atmospheric temperature, the main atmospheric component and the abundance of the trace gases. We  also explore the possible correlations between the said parameters.  To check the consistency of the solution retrieved and connect it to possible formation and evolution scenarios, we run in parallel an ab-initio chemical model developed by \cite{Venot2012} to study hot exoplanets' atmospheres. This model allows us to explore the impact of different elemental compositions -- in particular the hydrogen fraction and the C/O ratio -- on the molecular composition of the atmosphere. 

\section{DATA  ANALYSIS} \label{sec:analysis}

\subsection{Observations} \label{sub:observation}

We downloaded the publicly available spectroscopic images of the transiting exoplanet 55 Cancri e (ID: 13665, PI: Bj{\"o}rn Benneke) from the MAST Archive. The dataset, obtained with the G141 grism, contains two visits of four HST orbits each. Each exposure is the result of three up-the-ramp samples with a size of $522 \times 522$ pixels in the SPARS10 mode, resulting in a total exposure time of 8.774724 seconds, a maximum signal level of $6.7 \times 10^4$ electrons per pixel and a total scanning length of approximately 350 pixels. Both scanning directions were used for these observations, and in the text we will refer to them as forward (increasing row number) and reverse  (decreasing row number) scans. In addition, the dataset contains, for calibration purposes, a non-dispersed (direct) image of the target with the F132N filter.

The configuration of this particular dataset allows us to fully capitalize on our WFC3 scanning-mode pipeline, as the long scanning length and the high scanning speed intensify the effects of the geometrical distortions, as discussed in \cite{Tsiaras2015}. In addition we can study the coupling between the scanning speed and the reading process which are critical elements in fast scanning observations (Section \ref{sub:up_down_stream}). 

\subsection{Extracting the spectro-photometric light-curves} \label{sub:extraction}

In this section we summarize the steps to extract the spectro-photometric light-curves from the spatially scanned spectra. We refer the reader to \cite{Tsiaras2015} for a detailed discussion of our pipeline, which was conceived to reduce spatially scanned spectra. 

The first step is the standard reduction of the raw scientific frames: bias drifts correction, zero read subtraction, non-linearity correction, dark current subtraction, gain variations calibration, sky background subtraction and finally bad pixels and cosmic rays correction.

The next, fundamental step, is to calibrate the position of the frames as this is required for wavelength calibration. In scanning-mode observations, position shifts are common. We measure the horizontal shifts by comparing the normalized sum across the columns of each frame to the first frame of the first visit. The forward and reverse scans are compared to the first forward or reverse scanned spectrum respectively. This separation is necessary because the frames resulting from the two different scanning directions have slightly different position on the detector and consequently slightly different structure, which could bias our measurement. For the vertical shifts, we make use of the first non-destructive read of each exposure. We fit an extended Gaussian function on each column profile because, due to the very fast scanning, even these spectra are elongated. Finally, we fit the central values of these extended Gaussians as a linear function of column number, which indicate any possible vertical shifts. Again, we keep forward and reverse scans separated as their initial scanning positions are close to the two vertically opposite edges of the frame. 

\begin{figure}
	\centering
	\includegraphics[width=\columnwidth]{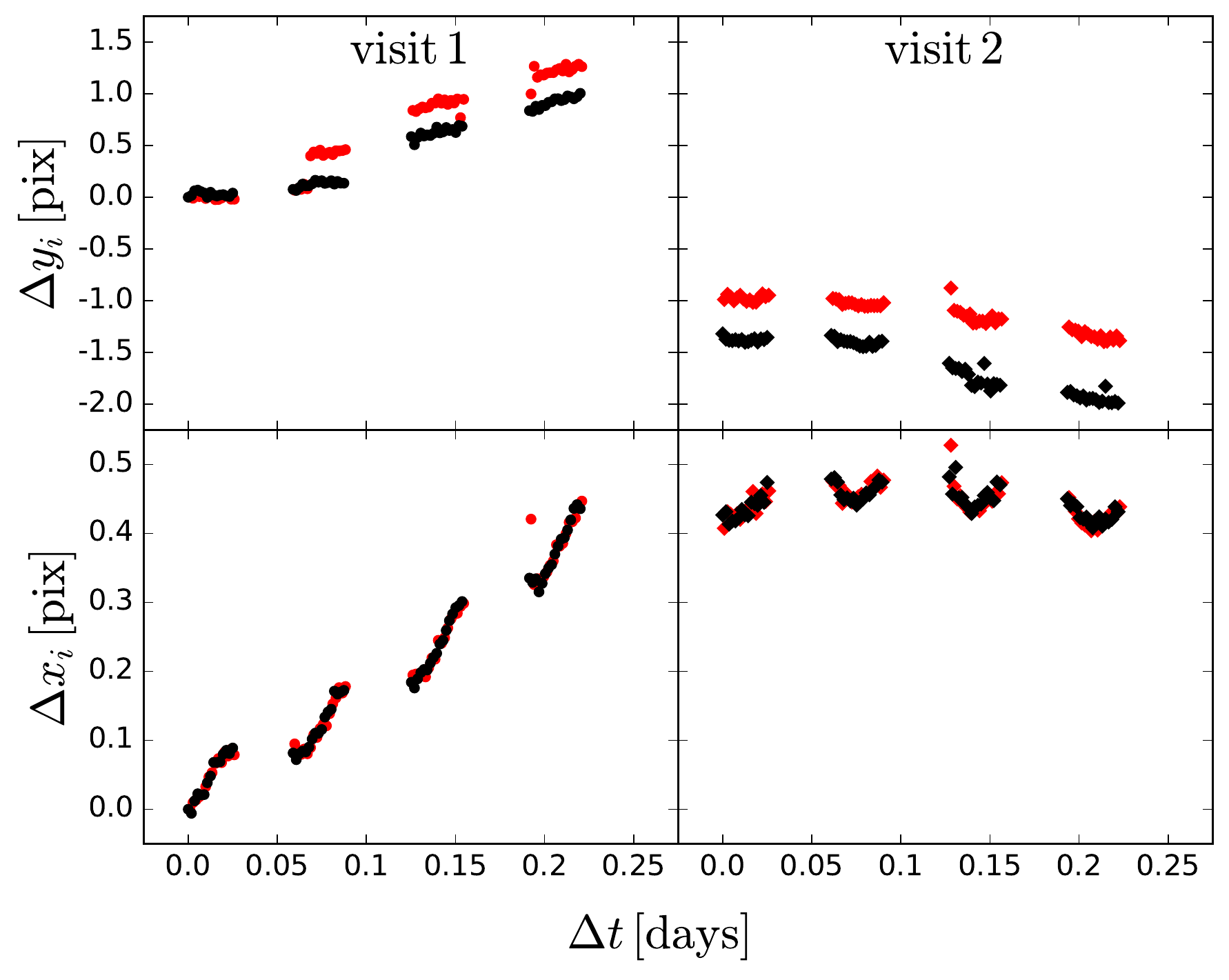}
	\caption{Vertical (top) and horizontal (bottom) shift for each frame of the two visits. The forward scans (black) are always compared to the first forward scan and the reverse scans (red) to the first reverse scan of the first visit.}
	\label{fig:shifts}
\end{figure}

In the first visit we find that the vertical shifts have an amplitude three times larger than the horizontal ones, while in the second visit they are five times larger (Figure \ref{fig:shifts}). The behaviour of the 55 Cancri e dataset is not commonly observed and a plausible explanation is the high scanning speed used.

We then proceed to the wavelength calibration. Based on the direct image included in the dataset, the calibration coefficients provided by \cite{coefficients} and the measured horizontal shifts, we calculate the wavelength-dependent photon trajectories for each frame. The wavelength-dependent photon trajectories indicate the position of photons at each wavelength on the detector during the scan. 

\begin{figure}
	\centering
	\includegraphics[width=\columnwidth]{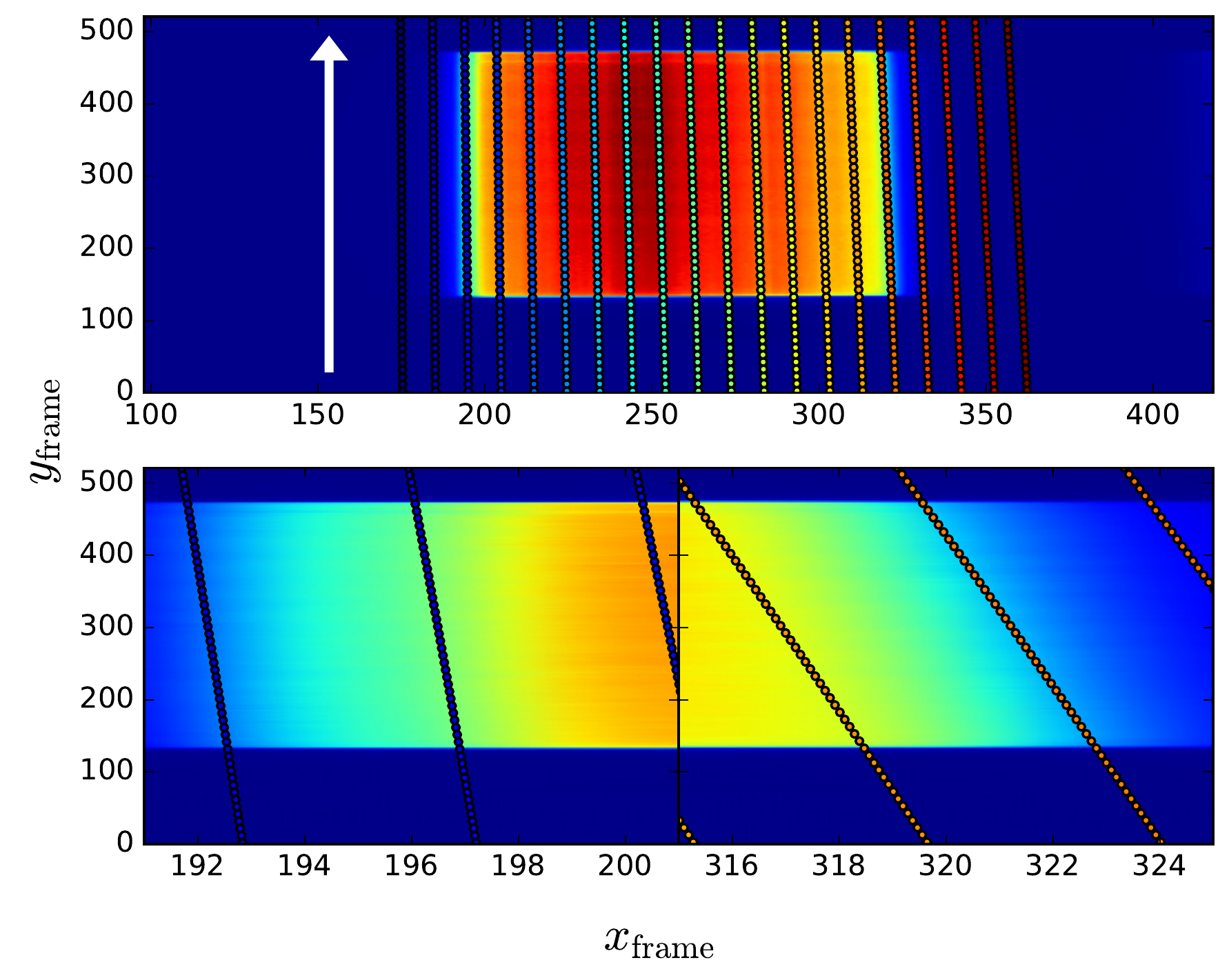}
	\caption{Top: Wavelength-dependent photon trajectories (colored points) and the trajectory of the relative direct image (white arrow). Bottom: Left and right edges of the spectrum where we can see the large difference between the position of photons with the same wavelength at the bottom and the top of the frame.}
	\label{fig:wst}
\end{figure}

As we can see in Figure \ref{fig:wst}, the position of photons at a particular wavelength are shifted towards the left part of the detector, as the scan processes. This behavior is caused by dispersion variations that our approach takes into account accurately. The effect on a single spatially scanned spectrum is that a pixel at the bottom of a given column is probing a different part of the stellar spectrum compared to a pixel at the top of the same column. For a column at $\sim1.2 \, \mu m$, the wavelength difference between the lower and the upper edge of the spatially scanned spectrum is $62 \, \AA$, 33\% of the total bin size of $185 \, \AA$. The effect is stronger at longer wavelengths ($\sim1.6 \, \mu m$), where the difference is $140 \, \AA$, 57\% of the total bin size of $230 \, \AA$. Given that the wavelength range inside one pixel is about $46 \AA$, in the first case the shift corresponds to 1.3 pixels, while in the second case, to 3 pixels. We plan to further investigate how this bias propagates to the final planetary spectrum by using our WFC3 simulator, \textit{Wayne} \citep{Varley2015}, in a future work.

\begin{figure}
	\centering
	\includegraphics[width=\columnwidth]{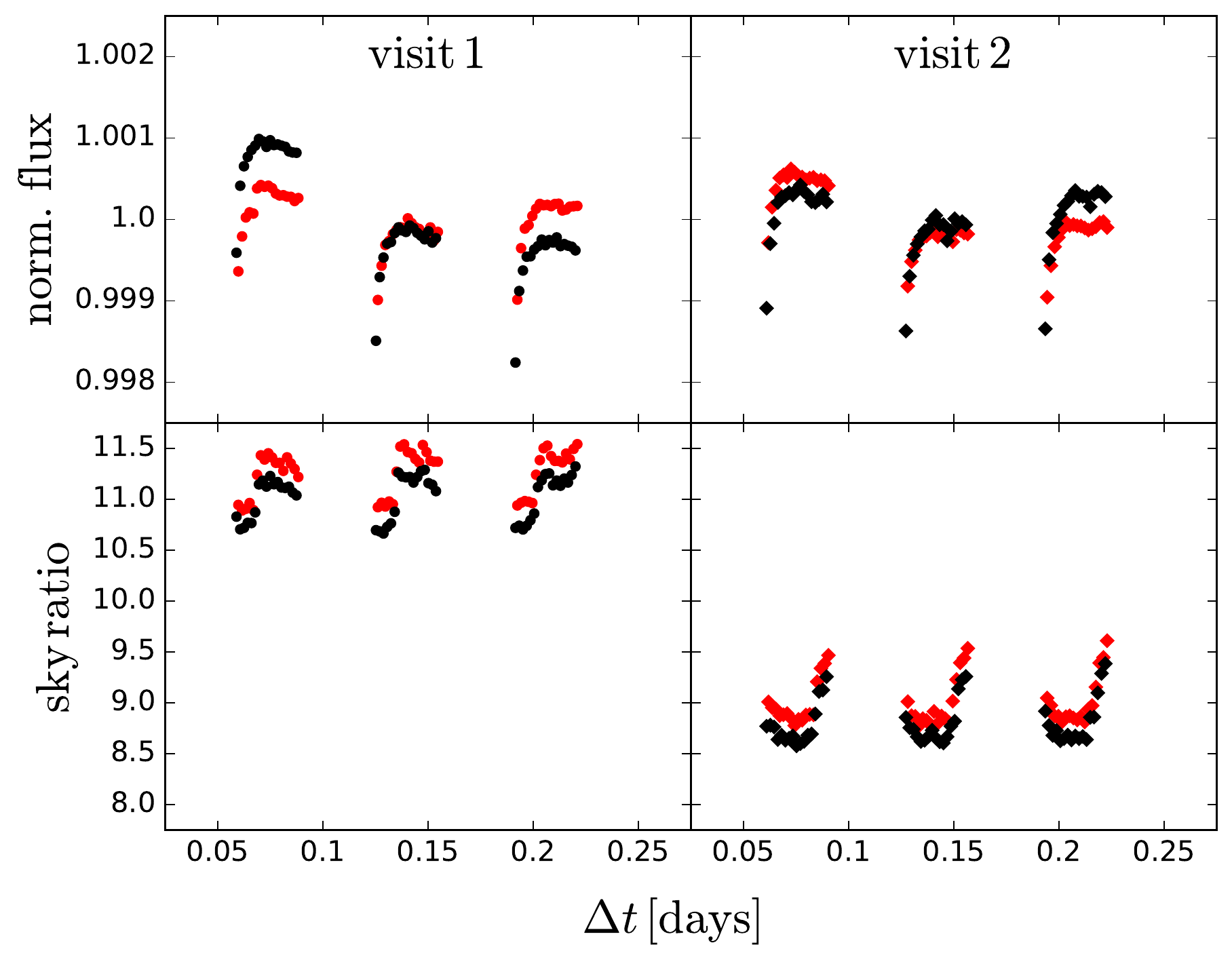}
	\caption{Top: Raw white light-curve for the forward (black) and reverse (red) scans, excluding the first orbit from each visit and normalized by the mean of each scanning direction for clarity. Bottom: Sky background relatively to the master sky frame for the two scanning directions in each visit.}
	\label{fig:rawlc}
\end{figure}

The photometric apertures are delimited by the wavelength-dependent photon trajectories of a selected binning grid and, as a result, they are of quadrangular shape. We take into account fractional pixels by using a second-order 2D polynomial distribution of the flux (based on the values of the surrounding pixels) and calculating the integral of this function inside the photometric aperture. By applying this process to all the frames we extract the white (Figure \ref{fig:rawlc}) and the spectral light-curves.

\subsection{Up-stream / Down-stream effect: exposure time \\ variations due to vertical position shifts} \label{sub:up_down_stream}

Reading the IR detector of the WFC3 camera takes about 2.93 s in total and, as a result, while the exposure time is the same for all the pixels, they are not all exposed simultaneously. The first and last rows of the detector are the first to be read, then the process continues to the inner rows until the vertical mid-point is reached. The reading direction is, therefore, different for the two upper and the two lower quadrants, as the first are read downwards and the second upwards. 

This process does not affect the staring-mode frames because the position of the spectrum is always the same. However, when the spatial scanning technique is used, there is a coupling between scanning and reading directions, referred to as down-stream/up-stream configuration, depending on whether the reading and scanning directions are the same/opposite. As a result, a down-stream configuration has a longer effective exposure time \citep{considerations}, which allows a longer total length for the scan and a higher total flux compared to an up-stream configuration.

\begin{figure}
	\centering
	\includegraphics[width=\columnwidth]{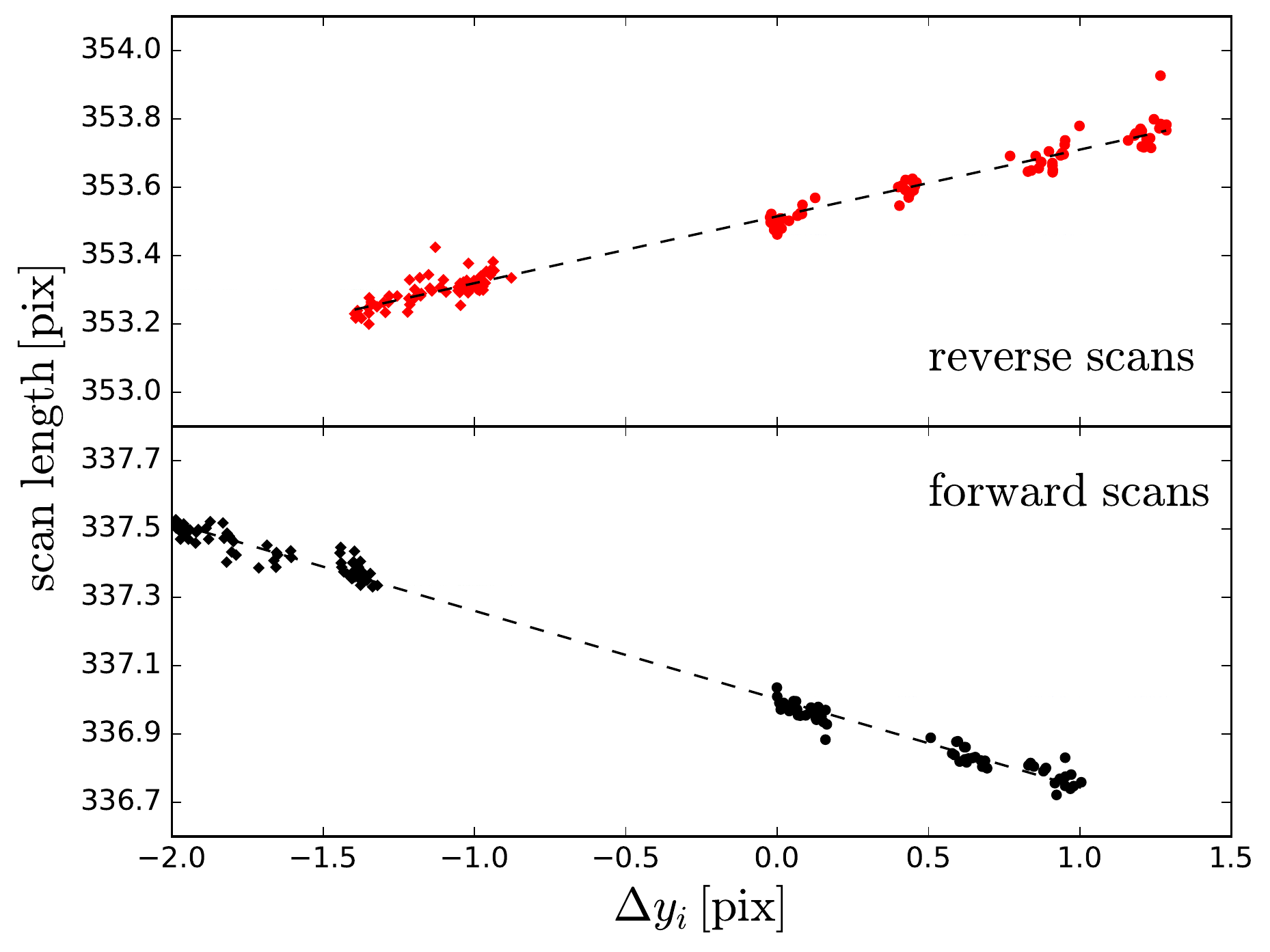}
	\caption{Scanning length variations with vertical position shifts for the two scanning directions in each visit (visit 1 - circles, visit 2 - diamonds)} and fitted models. The shifts are calculated with respect to first frame of the first visit for each scanning direction.
	\label{fig:stream}
\end{figure}

Figure \ref{fig:rawlc} shows clearly that the white light-curves of the two different scanning directions have different long-term slopes.  In particular, the reverse scans have an upwards trend relative to the forward scans in the first visit, and a downwards trend in the second visit. These trends are correlated with the vertical position shifts (Figure \ref{fig:shifts}). Indeed, we find that the total length of the scans -- i.e. the effective exposure time and therefore the total flux -- changes with the vertical position shifts, in a different way for the two scanning directions (Figure \ref{fig:stream}).

In cases where the spectrum trace crosses the mid-point line, the scanning/reading configuration is down-stream at the beginning of the scanning process and up-stream at the end of it. We believe that this effect could cause a dependency between the vertical position of the spectrum on the detector and the effective exposure time. For this dataset, the flux variation is about $0.7 \times 10^{-3}$ for a vertical shift of one pixel.

We correct for this effect using the linear behavior of the scanning length with the vertical position shifts (Figure \ref{fig:stream}) and rescaling each data point to the first of each scanning direction. Figure \ref{fig:fittingwl} plots both scanning directions which are fitted together with the same model, proving the consistency between them. Note that there is still a small, constant offset between them, so we have to fit for two different normalization factors.

\subsection{Fitting the white and the spectral light-curves} \label{sub:fitting}

The WFC3 infrared detector introduces time-dependent systematics, known as the ``ramps'', both in staring mode \citep{Berta2012, Swain2013, Wilkins2014} and scanning mode \citep{Deming2013, Kreidberg2014a, Knutson2014a, Tsiaras2015} observations. The effect of these systematics is stronger with increasing flux and, since 55 Cancri is a very bright source, the presence of the ``ramps'' is expected (Figure \ref{fig:rawlc}). We correct these systematics using an approach similar to \cite{Kreidberg2014a}. We adopt an instrumental systematics function:
\begin{equation}
	R(t) = (1 - r_a (t-t_\mathrm{v}))(1-r_{b1} e^{-r_{b2} (t-t_\mathrm{o})})
	\label{rampfunction}
\end{equation}

\noindent where $t$ is mid-time of each exposure, $t_\mathrm{v}$ the time when the visit starts, $t_\mathrm{o}$ the time when the orbit in which the frame belongs starts, $r_a$ the slop of the linear long-term ``ramp'' and ($r_{b1},r_{b2}$) the coefficients of the exponential short-term ``ramp''.

We use our transit light-curve model, which returns the relative flux, $F(t)$, as a function of time, given the limb darkening coefficients, $a_n$, the $R_\mathrm{p}/R_*$ ratio and all the orbital parameters ($T_0, P, i, a/R_*, e, \omega$). The model is based on the non-linear limb darkening law \citep{Claret2000} for the host star:
\begin{equation}
	I(a_n, r) =   1 - \sum_{n=1}^{n=4} a_n (1-(1-r^2)^{n/4})
	\label{limbdarkeninglaw}
\end{equation}

We derive the limb darkening coefficients from the \mbox{ATLAS} stellar model \citep{Kurucz1970, Howarth2011}, of a star similar to 55 Cancri (Table \ref{tab:parameters}), using the sensitivity curve of the G141 grism between 1.125 and 1.65  $\mu$m (Table \ref{tab:fitting}), to avoid its sharp edges. Since the ingress and egress of the transit are not included in the light-curve (Figure \ref{fig:rawlc}), it is impossible to fit for the orbital parameters. We therefore fix inclination, $a/R_*$ ratio and the mid-transit point to the values of Table \ref{tab:parameters}. We also assume a circular orbit.

\begin{figure}
	\centering
	\includegraphics[width=\columnwidth]{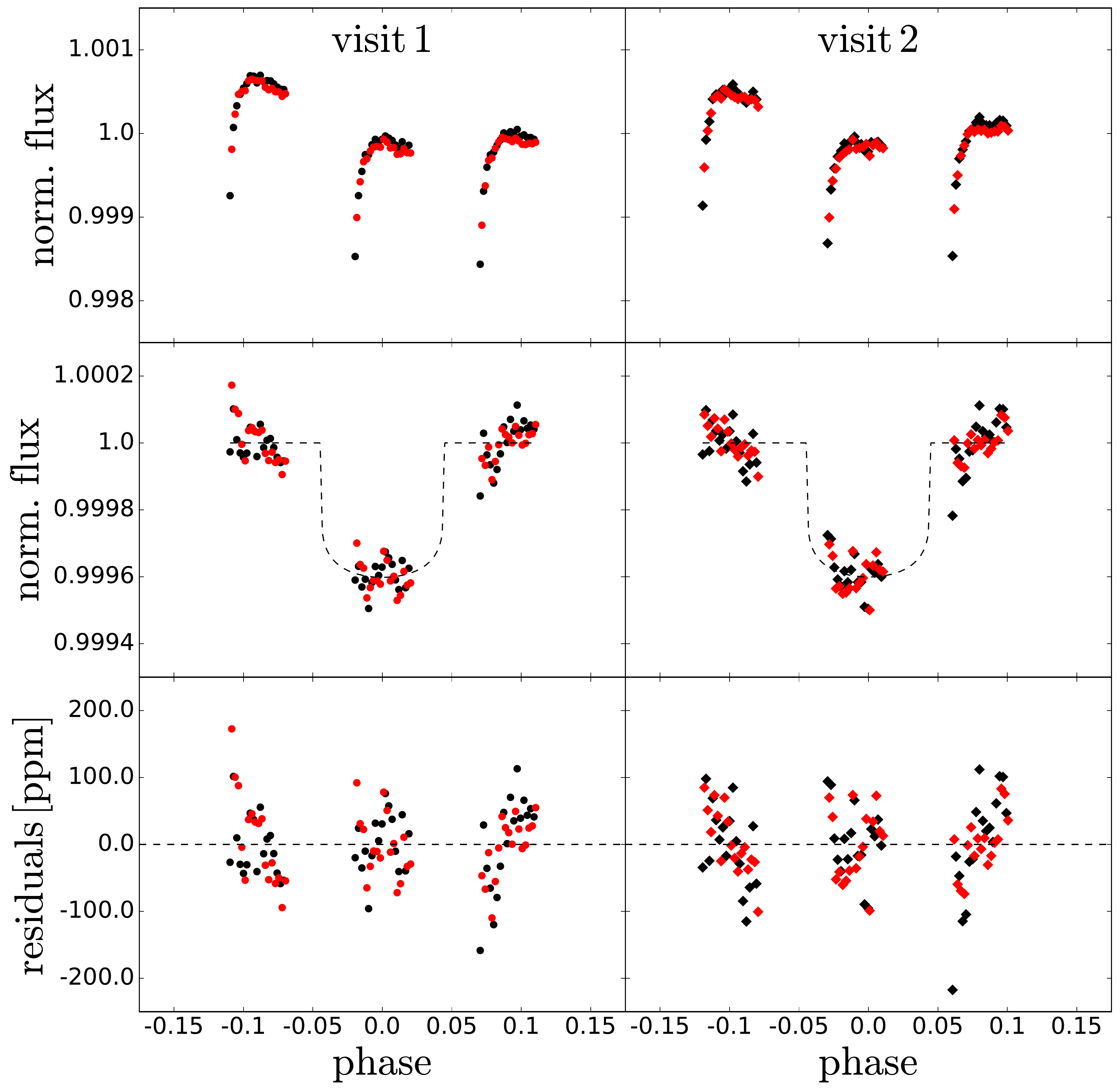}
	\caption{From top to bottom: 1) normalised raw light-curve for the forward (black) and reverse (red) scans, 2) normalised light-curve divided by the best fitted model for the systematics, compared to the fitted transit model and 3) fitting residuals.}
	\label{fig:fittingwl}
\end{figure}

\begin{figure}
	\centering
	\includegraphics[width=\columnwidth]{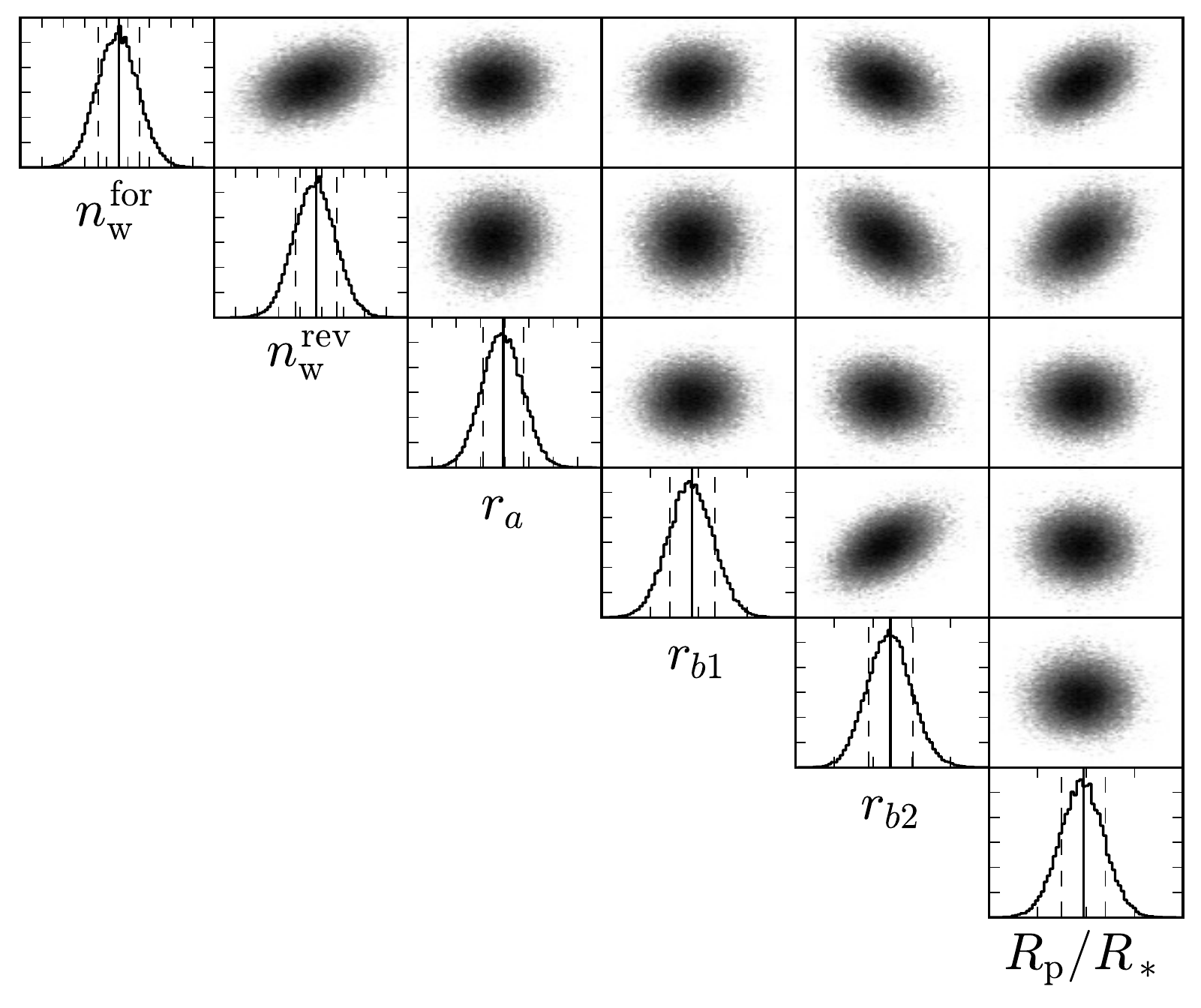}
	\caption{Correlations between the fitted systematics and transit parameters for the simultaneous fitting approach on the first visit data points. As we can see, apart from the expected correlation with the normalisation factors, $R_\mathrm{p}/R_*$ ratio is not correlated with any of the three parameters that describe the systematics.}
	\label{fig:correlations}
\end{figure}

The models of the systematics, $R(t)$, and the white transit light-curve, $F_\mathrm{w}(t)$, are both fitted to the white light-curve, following two different approaches:  a) fitting the transit model multiplied by the instrumental systematics and a normalization factor, $n_\mathrm{w}^\mathrm{scan} R(t) F_\mathrm{w}(t)$, and b) fitting the instrumental systematics, $R(t)$, on the out-of-transit points, correcting the light-curve, and then fitting a normalized transit model, $n_\mathrm{w}^\mathrm{scan} F_\mathrm{w}(t)$. The normalization factor, $n^\mathrm{scan}_\mathrm{w}$, changes to $n^\mathrm{for}_\mathrm{w}$ when the scanning direction is upwards and to $n^\mathrm{rev}_\mathrm{w}$when it is downwards, to account for the offset between them. Figure \ref{fig:fittingwl} shows only the case of simultaneous fitting as both approaches give the same result for each visit. The final $R_\mathrm{p}/R_*$ ratio is well consistent between the two visits (Table \ref{tab:fitting}). Figure \ref{fig:correlations} plots the correlations between the fitted parameters, showing no correlation between the $R_\mathrm{p}/R_*$ ratio and any of the parameters describing the systematics.

\begin{table}
	\small
	\center
	\caption{White light-curve fitting results.}
	\label{tab:fitting}
	\begin{tabular}{c | c}
		
		\hline \hline
		\multicolumn{2}{c}{Limb darkening coefficients (1.125 - 1.650 $\mu$m)}	\\ [0.1ex]
		\hline	
		$a_1$ 				& 0.695623							\\
		$a_2$				& $-$0.268246							\\
		$a_3$				& 0.357576							\\
		$a_4$				& $-$0.171548							\\ [1.0ex]
		
		\hline\hline
		\multicolumn{2}{c}{$R_\mathrm{p}/R_*$}							\\ [0.1ex]
		\hline
		visit 1	& 0.01894 $\pm$ 0.00023								\\
		visit 2	& 0.01892 $\pm$ 0.00023								\\
			
	\end{tabular}
\end{table}

As we can see in Figure \ref{fig:rawlc}, the sky background is slightly different for the forward and reverse scans, indicating that the area from which the sky background level is estimated is contaminated by stellar flux. Since the area we are using is already at the left edge of the detector, this contamination is inevitable. To measure its effect on the final results we repeat the whole process with an aperture which extends 10 pixels above and below the vertical edges of the spatially scanned spectrum (instead of the 20 used for the above results). We find a uniform shift of 3 ppm across the wavelength channels, while the uncertainties in the transit depth is, on average, 22 ppm for the spectral light-curves. This result indicates that the relative structure of the spectrum is not affected, and also the uncertainty in the $R_\mathrm{p}/R_*$ ratio includes this potential bias.

To calculate the wavelength-dependent transit depth we use the spectral light-curves divided by the white one. In this way we minimize the effect of the systematics model not perfectly approximating the real systematics, as the residuals in Figure \ref{fig:fittingwl} do not follow a gaussian distribution. The relative light-curves appear to have an out-of-transit slope which we interpret as an effect caused either by the horizontal shifts (model 1) or by a wavelength-dependent long-term ``ramp'' (model 2). In the first case we fit for a factor which is linear with the horizontal shifts, as calculated in Section \ref{sub:extraction}, while in the second case a factor which is linear with time. The models used for each of the above approaches are:
\begin{equation}
	\begin{split}
		\text{model 1: \, }	&	n^\mathrm{scan}_\lambda (1 + \chi_\lambda \Delta x) (F_\lambda/F_\mathrm{w}) \\ 
		\text{model 2: \, }	&	n^\mathrm{scan}_\lambda (1 + \chi_\lambda t) (F_\lambda/F_\mathrm{w})
	\end{split}
	\label{eq:sectral_ramp}
\end{equation}

\noindent where $n^\mathrm{scan}_\lambda$ is a normalisation factor with the same behaviour as in the white light-curve, $\chi_\lambda$ the linear factor in each case, $\Delta x = \Delta x(t)$ the horizontal shifts, $F_\mathrm{w}=F_\mathrm{w}(t)$ the white light-curve model and $F_\lambda=F_\lambda(t)$ the spectral light-curve model. Fitting the normalization factor per wavelength channel is necessary, as the offset between the scanning directions varies with wavelength. The orbital  parameters used are those from Table \ref{tab:fitting} and the white $R_\mathrm{p}/R_*$ ratio is taken from the fitting above.

For each spectral light-curve, we fit the above models twice: the first time, using the uncertainties provided by the reduction and extraction process and the second, using the standard deviation of the residuals of the first fit. The final uncertainty reported is the maximum of the two and it is between 5\% and 20\% above the photon noise limit. In this way we take into account possible scatter in the data points that is additional to the photon noise and the noise introduced by the instrument.

The wavelength bins are selected within the limits of the white light light-curve, and with constant resolution of $\lambda / \Delta \lambda = 65$, which is half of the resolving power of the G141 grism, to keep an approximately uniform SNR. We calculate the limb darkening coefficients for each spectral light-curve using the \mbox{ATLAS} model, the stellar parameters in Table \ref{tab:parameters} and the sensitivity curve of the G141 grism inside the boundaries of each wavelength bin.

\subsection{Atmospheric retrieval} \label{sub:retrieval}

In order to fit the WFC3 spectrum we use \taurex \citep{Waldmann2015a,Waldmann2015b}, a Bayesian spectral retrieval that fully maps the parameter space through the nested sampling algorithm. The transmission spectra are generated using cross sections based on the line lists from \mbox{ExoMol} \citep{jt528,jt378,jt500,jt563,jt564,jt570}, HITRAN \citep{HITRAN2009,HITRAN2013} and HITEMP \citep{HITEMP2010}. The atmosphere is parametrized assuming an isothermal profile with constant molecular abundances as a function of pressure. We include a wide range of molecules in the fit, including H$_2$O, HCN, NH$_3$, CH$_4$, CO$_2$, CO, NO, SiO, TiO, VO, H$_2$S, C$_2$H$_2$. The fitted parameters are the mixing ratios of these molecules, the atmospheric mean molecular weight, the surface pressure and radius. 

We use uniform priors for the gas mixing ratios ranging from 1 to 10$^{-8}$. The mean molecular weight is coupled to the fitted composition, in order to account for both the fitted trace gases and possible unseen absorbers with signatures outside the wavelength range probed here. The uniform prior assumed for the mean molecular weight ranges from 2 to 10 amu. Lastly, we assume uniform priors for the surface radius, the surface pressure and the mean atmospheric temperature, ranging between $0.1 - 0.3 \, R_\mathrm{Jup}$, $10-10^7 \, \mathrm{Pa}$, and $2100-2700 \, \mathrm{K}$ respectively. We do not include a separate parameterization for the cloud layer, the pressure at the surface could be the pressure at the top of a cloud deck. 

\subsection{Ab-initio chemical simulations} \label{sub:chemistry}

In parallel to the spectral retrieval, we investigated the theoretical predictions for the chemical composition of an atmosphere enveloping 55 Cancri e. We assumed an atmosphere dominated by hydrogen and helium, with a mean molecular weight of 2.3 amu. We used a thermal profile with high-altitude atmospheric temperature of 1600 K. These parameters correspond to a scale height of 440 km, or 25 ppm, at the surface.

We used the same chemical scheme implemented in \cite{Venot2012} to produce vertical abundance profiles for 55 Cancri e, assuming a solar C/O ratio and a C/O ratio of 1.1. This chemical scheme can describe the kinetics of species with up to two carbon atoms. In \cite{Venot2015} it was shown that a more sophisticated chemical scheme including species with up to six carbon atoms produces comparable results, and that the simpler scheme can reliably model atmospheres with C/O ratios above 1.  The scheme has been developed with combustion specialists and validated by experiments conducted in a wide range of temperatures ($300 - 2500 \, \mathrm{K}$) and pressures ($0.01 - 100 \, \mathrm{bar}$) \citep[e.g.][]{Battin2006, Bounaceur2007, Anderlohr2010, Bounaceur2010, Wang2010}.

The stellar flux was calculated in the following way. From 1 to 114 nm, we used the mean of the Sun spectra at maximum and minimum activity \citep{Gueymard2004}, scaled for the radius and effective temperature of 55 Cancri. From 115 to 900 nm, we used the stellar flux of $\varepsilon$ Eridani (HD22049) from \cite{Segura2003} scaled also to the properties of 55 Cancri. $\varepsilon$ Eridani is a K2V star ($T_\mathrm{eff} = 5084 \, \mathrm{K}$ and $R = 0.735 \, R_\mathrm{sun}$) quite close to 55 Cancri, which is a G8V star ($T_\mathrm{eff} = 5196 \, \mathrm{K}$ and $R = 0.943 \, R_\mathrm{sun}$), making $\varepsilon$ Eridani a quite good proxy for 55 Cancri.

\section{RESULTS} \label{sec:results}

\begin{figure}
	\centering
	\includegraphics[width=\columnwidth]{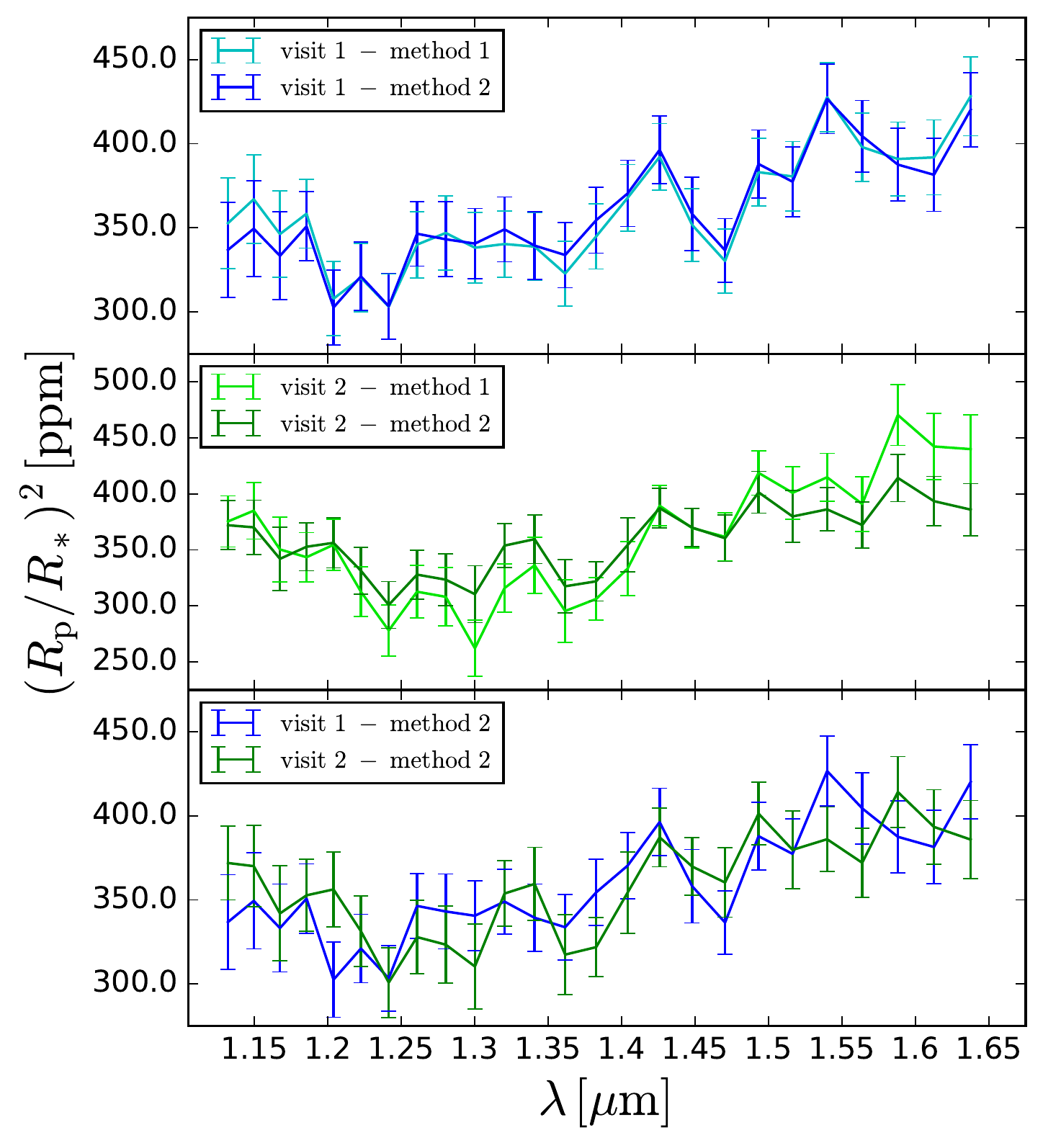}
	\caption{Results from fitting the spectral light-curves. From top to bottom: 1) comparison of the two models for the first visit, 2) comparison of the two models for the second visit, 3) comparison of model 2 for both visits.}
	\label{fig:spectrum}
\end{figure} 

\begin{table*}
	\small
	\center
	\caption{Limb darkening coefficients $a_{1-4}$ and transit depth $(R_\mathrm{p}/R_*)^2$ for the wavelength channels.}
	\label{tab:spectrum}
	\begin{tabular}{r l | c c c c | c c c}
		\hline \hline
				&									&			 &			 &			 &			 & \multicolumn{3}{| c}{$(R_\mathrm{p}/R_*)^2 \, \mathrm{[ppm]}$}		\\
		\multicolumn{2}{c |}{$\lambda_1- \lambda_2 \, [\mu \mathrm{m}]$} 	& $a_1$ 		 & $a_2$ 		 & $a_3$ 		 & $a_4$ 		 &  visit 1		& visit 2		& w. average			\\ [0.1ex]
		\hline
		1.1233 	 & 1.1407 								 & 0.772891 	 & $-$0.719847 	 & 0.973720 	 & $-$0.395615 	 & 337 $\pm$ 28	& 372 $\pm$ 23	& 358 $\pm$ 18	\\ 
		1.1407 	 & 1.1584 								 & 0.748548 	 & $-$0.643287 	 & 0.890964 	 & $-$0.366607 	 & 349 $\pm$ 29	& 370 $\pm$ 24	& 361 $\pm$ 18	\\ 
		1.1584 	 & 1.1764 								 & 0.738983 	 & $-$0.606810 	 & 0.847497 	 & $-$0.353214 	 & 333 $\pm$ 26	& 342 $\pm$ 28	& 337 $\pm$ 19	\\ 
		1.1764 	 & 1.1946 								 & 0.729208 	 & $-$0.572574 	 & 0.800655 	 & $-$0.335301 	 & 351 $\pm$ 21	& 353 $\pm$ 22	& 352 $\pm$ 15	\\ 
		1.1946 	 & 1.2131 								 & 0.725726 	 & $-$0.564131 	 & 0.785763 	 & $-$0.330767 	 & 303 $\pm$ 22	& 357 $\pm$ 22	& 329 $\pm$ 16	\\ 
		1.2131 	 & 1.2319 								 & 0.697907 	 & $-$0.466097 	 & 0.687658 	 & $-$0.298018 	 & 321 $\pm$ 20	& 332 $\pm$ 21	& 326 $\pm$ 14	\\ 
		1.2319 	 & 1.2510 								 & 0.692692 	 & $-$0.446746 	 & 0.662990 	 & $-$0.289484 	 & 303 $\pm$ 20	& 301 $\pm$ 21	& 302 $\pm$ 14	\\ 
		1.2510 	 & 1.2704 								 & 0.685452 	 & $-$0.409084 	 & 0.612133 	 & $-$0.271088 	 & 346 $\pm$ 20	& 328 $\pm$ 22	& 338 $\pm$ 15	\\ 
		1.2704 	 & 1.2901 								 & 0.685892 	 & $-$0.378366 	 & 0.574713 	 & $-$0.270881 	 & 343 $\pm$ 22	& 324 $\pm$ 23	& 333 $\pm$ 16	\\ 
		1.2901 	 & 1.3101 								 & 0.673686 	 & $-$0.354792 	 & 0.546395 	 & $-$0.250464 	 & 341 $\pm$ 21	& 311 $\pm$ 25		& 328 $\pm$ 16	\\ 
		1.3101 	 & 1.3304 								 & 0.669421 	 & $-$0.331802 	 & 0.512473 	 & $-$0.238250 	 & 349 $\pm$ 20	& 354 $\pm$ 20	& 351 $\pm$ 14	\\ 
		1.3304 	 & 1.3511 								 & 0.660777 	 & $-$0.276281 	 & 0.439988 	 & $-$0.211147 		 & 339 $\pm$ 20	& 360 $\pm$ 22	& 348 $\pm$ 15	\\ 
		1.3511 	 & 1.3720 								 & 0.659749 	 & $-$0.252857 	 & 0.401237 	 & $-$0.196660 	 & 334 $\pm$ 20	& 318 $\pm$ 24	& 327 $\pm$ 15	\\ 
		1.3720 	 & 1.3933 								 & 0.652560 	 & $-$0.189532 	 & 0.311675 	 & $-$0.162120 	 & 354 $\pm$ 20	& 322 $\pm$ 20	& 338 $\pm$ 14	\\ 
		1.3933 	 & 1.4149 								 & 0.650933 	 & $-$0.175302 	 & 0.300256 	 & $-$0.163030 	 & 370 $\pm$ 20	& 355 $\pm$ 24	& 363 $\pm$ 15	\\ 
		1.4149 	 & 1.4368 								 & 0.650697 	 & $-$0.189841 	 & 0.330447 	 & $-$0.180237 	 & 396 $\pm$ 20	& 387 $\pm$ 19	& 391 $\pm$ 14	\\ 
		1.4368 	 & 1.4591 								 & 0.650529 	 & $-$0.135161 	 & 0.234549 	 & $-$0.139640 	 & 358 $\pm$ 22	& 370 $\pm$ 19	& 365 $\pm$ 14	\\ 
		1.4591 	 & 1.4817 								 & 0.639029 	 & $-$0.076708 	 & 0.161044 	 & $-$0.112436 	 & 337 $\pm$ 20	& 361 $\pm$ 21	& 348 $\pm$ 14	\\ 
		1.4817 	 & 1.5047 								 & 0.655337 	 & $-$0.083499 	 & 0.125497 	 & $-$0.091207 	 & 388 $\pm$ 20	& 402 $\pm$ 20	& 394 $\pm$ 14	\\ 
		1.5047 	 & 1.5280 								 & 0.660770 	 & $-$0.025056 	 & 0.023486 	 & $-$0.048336 	 & 377 $\pm$ 21	& 380 $\pm$ 23	& 378 $\pm$ 16	\\ 
		1.5280 	 & 1.5517 								 & 0.686804 	 & $-$0.012316 	 & $-$0.049479	 & $-$0.011120 		 & 427 $\pm$ 21	& 386 $\pm$ 20	& 405 $\pm$ 14	\\ 
		1.5517 	 & 1.5758 								 & 0.721349 	 & $-$0.080742 	 & $-$0.011227	 & $-$0.017553 	 & 404 $\pm$ 21	& 372 $\pm$ 21	& 388 $\pm$ 15	\\ 
		1.5758 	 & 1.6002 								 & 0.756804 	 & $-$0.185811 	 & 0.070349 	 & $-$0.038516 	 & 388 $\pm$ 22	& 414 $\pm$ 21	& 402 $\pm$ 15	\\ 
		1.6002 	 & 1.6250 								 & 0.801062 	 & $-$0.222450 	 & 0.054906 	 & $-$0.024391 	 & 382 $\pm$ 23 	& 394 $\pm$ 22	& 388 $\pm$ 16	\\  
		1.6250 	 & 1.6502 								 & 0.828396 	 & $-$0.271626 	 & 0.075149 	 & $-$0.025545 	 & 420 $\pm$ 23 	& 386 $\pm$ 23	& 404 $\pm$ 16	\\  			
	\end{tabular}
\end{table*}

\begin{figure*}
	\centering
	\includegraphics[width=0.69\textwidth]{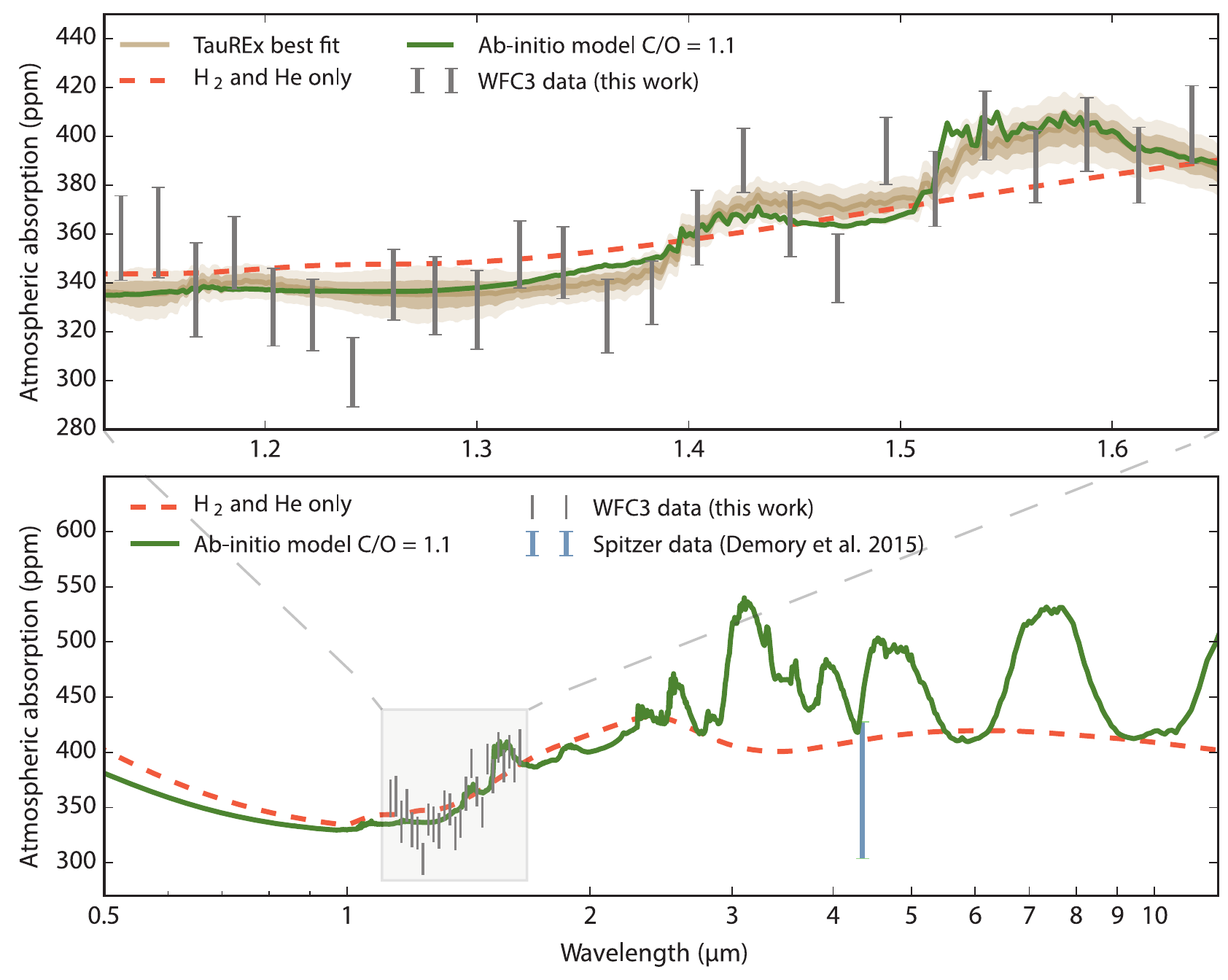}
	\caption{Top: Infrared transmission spectrum of the hot super-Earth 55 Cancri e (grey error bars), best fit obtained with \taurex (brown line), fitted model containing hydrogen and helium (dashed orange line) and an ab-initio model with C/O = 1.1 (green line, see also Section \ref{sec:hcnco}). The shaded regions show the the 1 and 2$\sigma$ confidence intervals in the the retrieved spectrum. Bottom: The same hydrogen/helium and ab-initio models plotted in a broader wavelength range. As we can see the two models can be better distinguished at longer wavelengths. The average transit depth of 55 Cancri e at 4.5 mu obtained with Spitzer Space Telescope \citep{Demory2015} is also shown in light blue.}
	\label{fig:spectrumfit}
\end{figure*}

\begin{figure}
	\centering
	\includegraphics[width=\columnwidth]{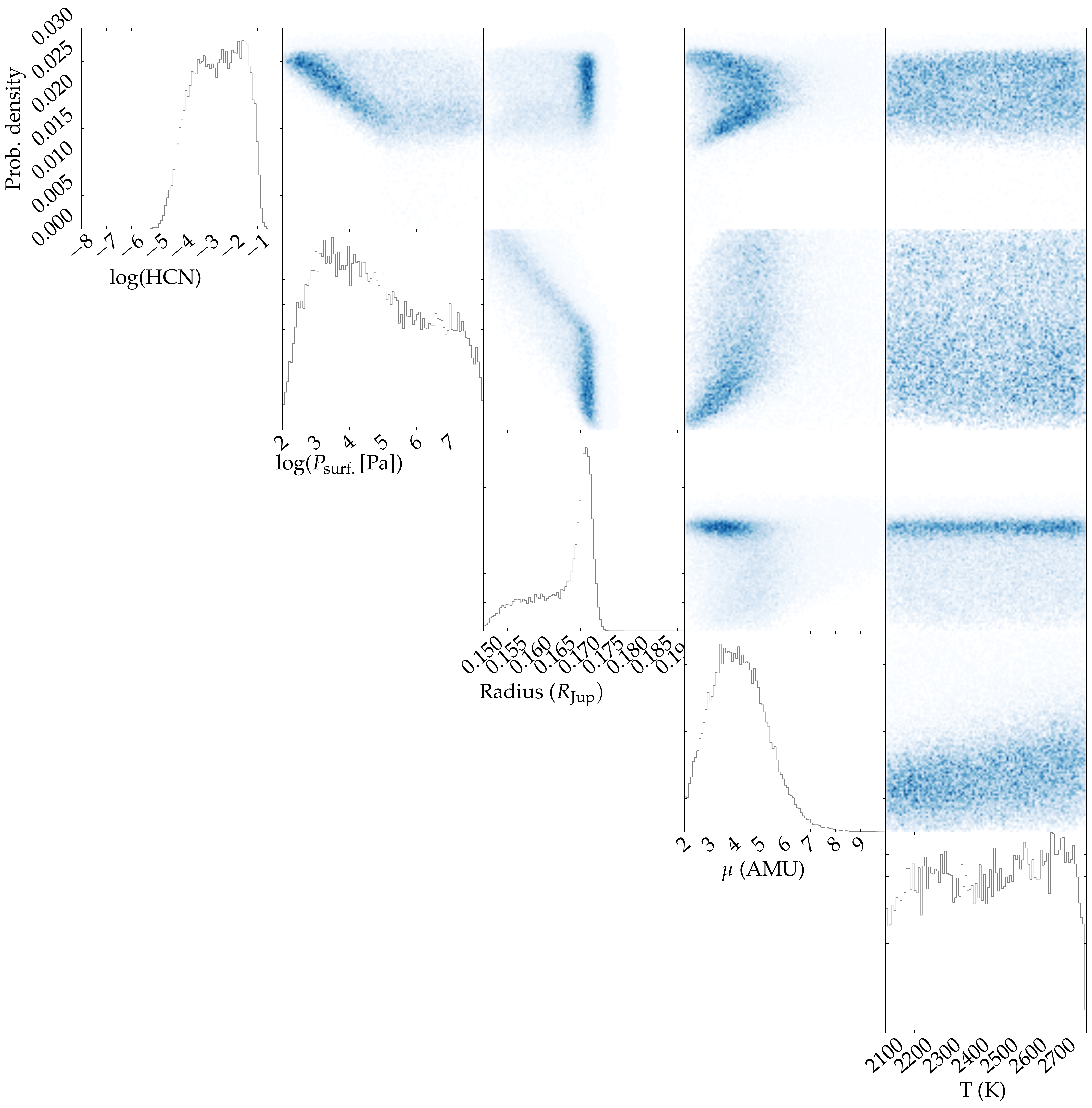}
	\caption{Posterior distributions of the retrieved atmospheric parameters and trace gases. Amongst all the molecules considered in the fit, here we only show the posterior of HCN, as all the other molecules show little or no contribution to the spectrum.}
	\label{fig:posterior}
\end{figure}

The results obtained with the two different fitting models described in Section \ref{sub:fitting} are plotted in Figure \ref{fig:spectrum}. In the first visit the transit depths are consistent within 0.3$\sigma$ and there is no preference between the two models, as fitting model 2 only improves the standard deviation of the residuals by 0.5\% compared to model 1. In the second visit, the results are less consistent with each other (1.1$\sigma$) because the horizontal shifts have a very small amplitude and cannot explain the observed slope. In this case, fitting model 2 improves the standard deviation of the residuals by 10\% and, therefore the final results that we report have been obtained with model 2. The calculated limb darkening coefficients ($a_{1-4}$) and the final measurements of the transit depths $(R_\mathrm{p}/R_*)^2$ as a function of wavelength are tabulated in Table \ref{tab:spectrum}. The transmission spectrum of 55 Cancri e and the best fits to it, obtained with \taurexng, are shown in Figure \ref{fig:spectrumfit}, while Figure \ref{fig:posterior} shows the posterior distributions of the fit to the spectrum, using the model with HCN.

Regardless of the specific gas causing the absorption features on the right hand side of the WFC3 spectrum, the mean molecular weight ($\mu$) of the atmosphere peaks at about 4 amu, as shown by the posterior distribution in Figure \ref{fig:posterior}. Higher values for the mean molecular weight would make the atmosphere more compact, and the features weakened. The relatively strong absorption seen between 1.4 and 1.6 $\mu$m indicates that $\mu$ is relatively low, and that the atmosphere is likely dominated by a mixture of hydrogen and helium. Therefore, the spectral absorbing features seen in the spectrum should be attributed to trace gases, rather than the main atmospheric component. The posterior of the surface pressure peaks at 0.1 bar, but we note that the parameter is only loosely constrained by the data.

\begin{figure}
	\centering
	\includegraphics[width=\columnwidth]{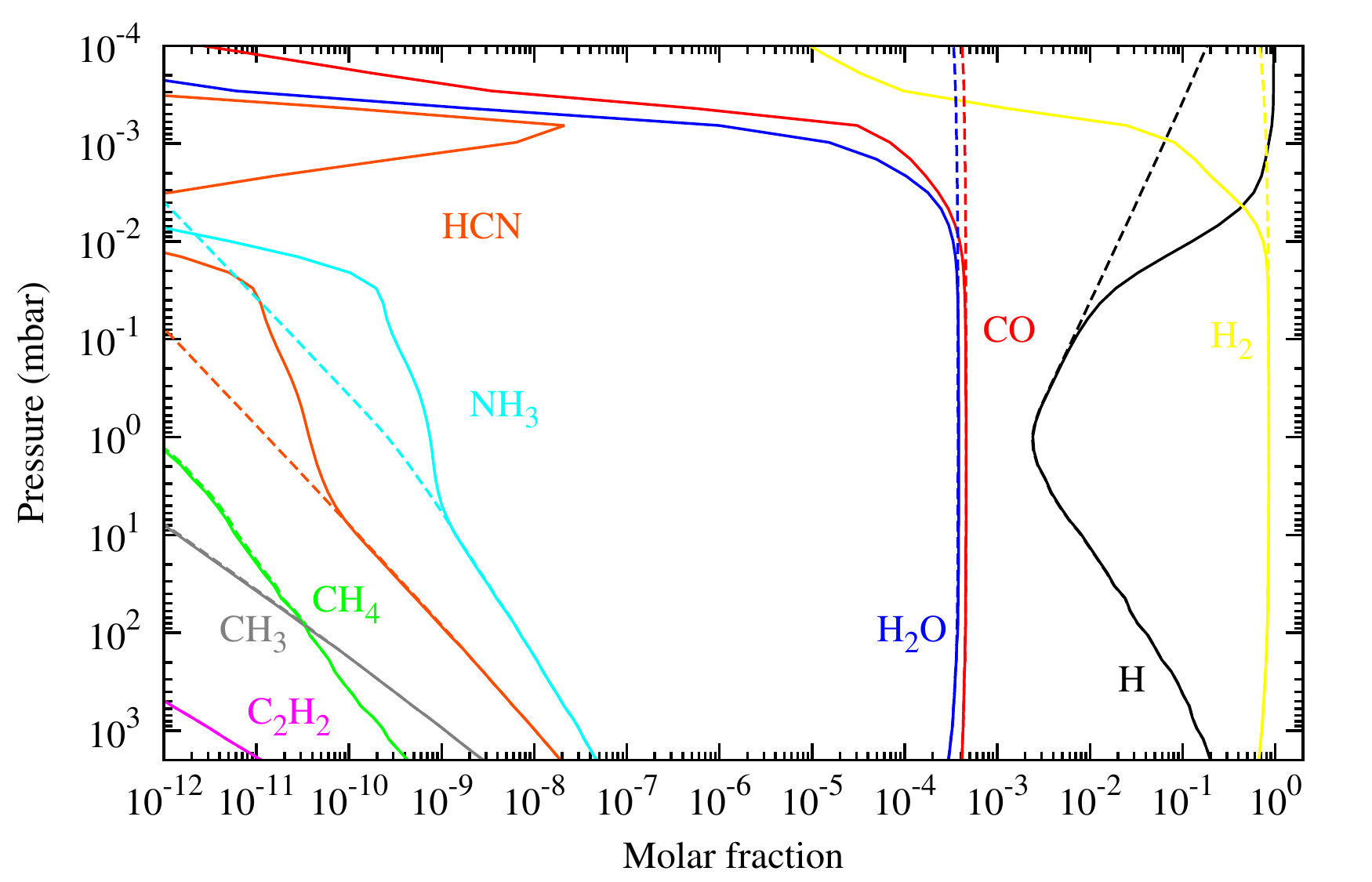}
	\includegraphics[width=\columnwidth]{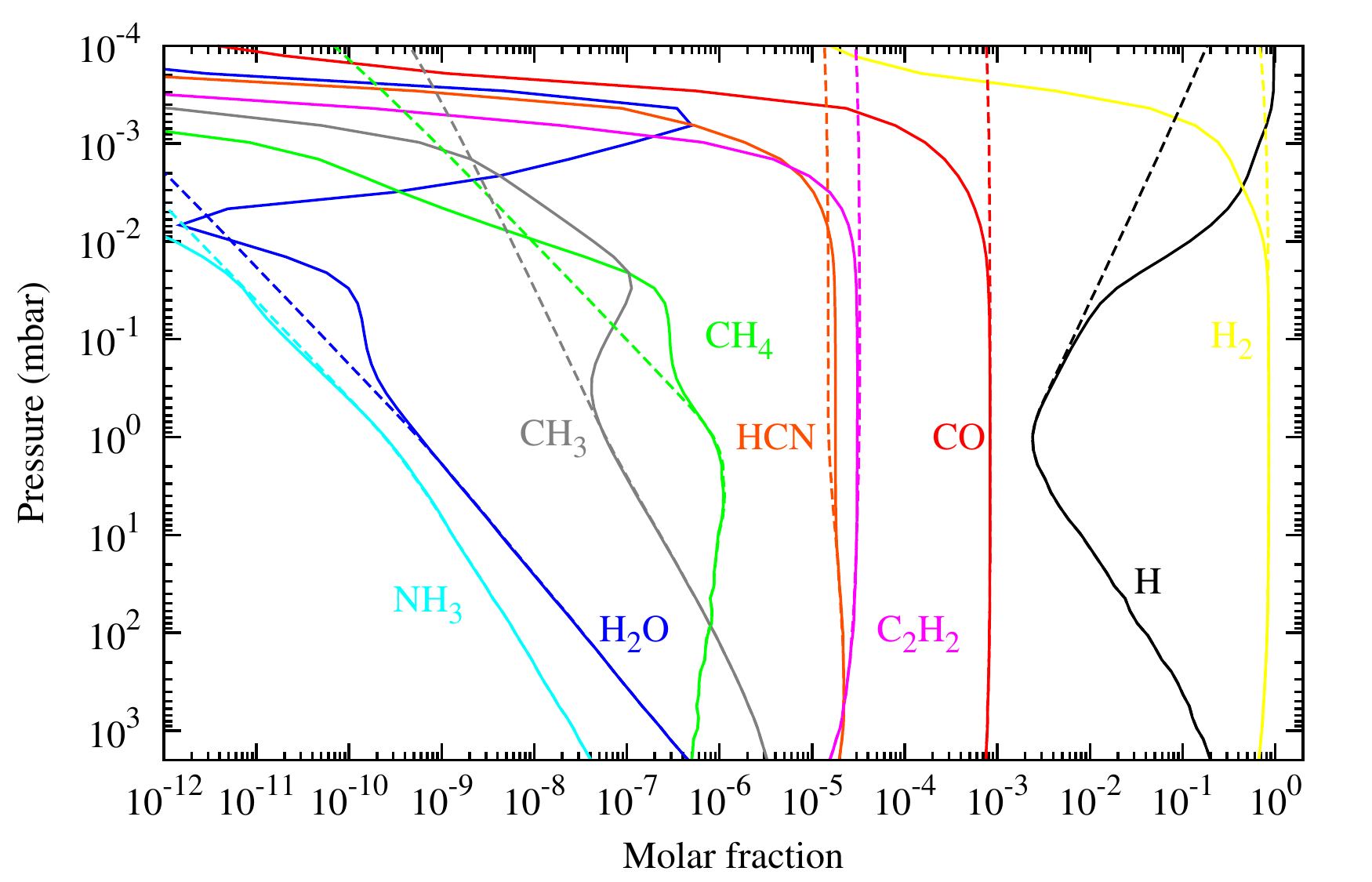}
	\caption{Vertical abundance profiles for the scenario with C/O solar (top) and C/O = 1.1 (bottom). The chemical compositions are calculated with the chemical scheme presented in \citep{Venot2012}, and includes the effects of photochemistry. The chemical equilibrium is also represented (dashed lines).}
	\label{fig:molprofiles}
\end{figure}

Amongst the molecules considered in the fit, we find that the best absorber that can fit the data is HCN. This species can adequately fit the absorption features seen at approximately 1.42 and 1.54 $\mu$m. All the other absorbers show little or no contribution to the overall spectrum. Interestingly, we find no evidence of water vapor. Despite the result of the retrieval, we stress that with the current precision of the measurements and this restricted wavelength range, we cannot confirm the presence or absence of certain absorbers, in particular CO$_2$, C$_2$H$_2$ and CO.

We find that the posterior distribution of HCN tends to favour a scenario with a high absolute mixing ratio, but the acceptable values are broad, starting at $10^{-5}$. We note the degeneracy between the HCN mixing ratio and the surface pressure: the lower the HCN abundance, the higher the surface pressure. Lastly, we find that the temperature is very poorly constrained by this data. For our best fitted model, which includes HCN, the scale height of the atmosphere is 242 km, or 14 ppm (with $T = 2100 \, \mathrm{K}$, $\mu = 3.2 \, \mathrm{amu}$, $R_\mathrm{p} = 0.17 \, R_\mathrm{Jup}$). Given this value, the amplitude of the spectral modulation of 70 ppm, corresponds to 5 scale heights.

We also run other models to verify the validity of this result. We find that a straight line fit has a $\chi^2 = 89.4$, which, with 24 degrees of freedom, indicates that a straight line can be rejected with a 6$\sigma$ confidence level. We also try to fit a model containing only a mixture of hydrogen and helium and no other trace gases. This model, which shows only the H$_2$-H$_2$ and H$_2$-He collision-induced absorption (dashed line in Figure \ref{fig:spectrumfit}) has a $\chi^2 = 39.6$.

As the nested sampling algorithm implemented in \taurex allows the precise computation of the global evidence of each model, we can also perform model comparison in a Bayesian framework. Table \ref{tab:models} summarizes the global evidence and $\chi^2$ values of the different models. We find that the straight line model has a log-evidence of 206.8, the model containing hydrogen and helium has $\log E = 226.2$, and the model containing HCN has $\log E = 228.8$. The Bayes factor $B_m$, defined as the ratio between the evidences of two different models, shows that we can confidently reject the straight line model. We find that the Bayes factors for the models with HCN or hydrogen and helium are 20 times larger compared to the straight line model, suggesting very strong preference for the former two models according to the Jeffreys' scale (Jeffreys 1961). Finally, the Bayes factor for the model with HCN is 2.6 times larger compared to the model with hydrogen and helium, suggesting a moderate preference for the model with HCN.

Interestingly, we find that the spectrum obtained using the results of the ab-initio model with a C/O greater than one is very close to the best fit found with \taurex (see Figure \ref{fig:spectrumfit}). In this case we obtain a $\chi^2$ of 26.1. Figure \ref{fig:molprofiles} shows the vertical abundance profiles for the two cases with C/O ratio solar (top) and C/O = 1.1 (bottom). It can be clearly seen that the two scenarios are significantly different. For the solar C/O case, the dominant absorbing gases are CO and H$_2$O, with mixing ratios of $\approx$ $3\times10^{-4}$ and $4\times10^{-4}$ respectively. On the contrary, for C/O = 1.1, while CO still remains the dominant species at $10^{-3}$, H$_2$O decreases to $10^{-7} -10^{-8}$, and HCN and C$_2$H$_2$ increase to about $10^{-5}$.

The absolute abundance of HCN expected for a C/O = 1.1 scenario is roughly consistent with what we found in the retrieved spectrum, being at the end of the left tail of the posterior distribution of the fit (Figure \ref{fig:posterior}). The transmission spectrum obtained using the abundances profiles for the C/O = 1.1 ratio is shown in Figure \ref{fig:spectrumfit} (green line), and shows that is roughly consistent with the retrieved spectrum. The dominant absorbers in this wavelength range seen in this ab-initio model is HCN, while other relatively strong absorbers, such as CO, C$_2$H$_2$ and CH$_4$, are all hidden below the HCN absorption.

\begin{table}
	\small
	\center
	\caption{Log-evidence ($\log E $) and $\chi^2$ values for the different models shown in Figure \ref{fig:spectrumfit}.}
	\label{tab:models}
	\begin{tabular}{l c c c}
		\hline \hline
 		Model					& $\log E $	& $\chi^2$		\\ [0.1ex]
		\hline
		Straight line				& 206.8		& 89.4		\\
		Helium and hydrogen only		& 226.2		& 39.6		\\       
		Best fit (including HCN)		& 228.8		& 23.7		\\
		Ab-initio model with C/O = 1.1	& --			& 26.1		\\
								&			&
       \end{tabular}
\end{table}

\section{DISCUSSION} \label{sec:discussion}

\subsection{Atmospheric features in the atmosphere \\ of 55 Cancri e}

The absorbing features seen in the spectrum of 55 Cancri e are indicative of the presence of an atmosphere. Our spectral retrieval shows that the atmosphere is likely dominated by a mixture of hydrogen and helium, and suggests that the features seen are mainly due to HCN. No signature of water vapor is found.

\subsection{HCN as a tracer of high C/O ratio atmospheres} \label{sec:hcnco}

If the features seen at 1.42 and 1.54 $\mu$m are due to hydrogen cyanide (HCN), the implications for the chemistry of 55 Cancri e are considerable. \cite{Venot2015}, using a new chemical scheme adapted to carbon-rich atmospheres, pointed out that the C/O ratio has a large influence on the C$_2$H$_2$ and HCN content in the exoplanet atmosphere, and that  C$_2$H$_2$ and HCN can act as tracers of the C/O ratio. Indeed, in a large range of temperatures above 1000 K, at a transition threshold of about C/O = 0.9, the C$_2$H$_2$ and HCN abundance increases by several orders of magnitude, while the H$_2$O abundance decreases drastically, as sown in Figure \ref{fig:molprofiles}. 

We conclude that if the absorption feature is confirmed to be due to HCN,  the implications are that this atmosphere has C/O ratio higher than solar.  However, additional data in a broader spectral range are necessary to confirm this scenario. We also note that while there is a good line list available for the hot HCN/HNC system \citep{jt298, jt570}, there is no comprehensive line list available for hot C$_2$H$_2$; provision of such a list is important for future studies of this interesting system.

 \section{CONCLUSIONS} 
 
In this paper we presented the first analysis of the two HST/WFC3 scanning-mode spectroscopic observations of the super-Earth 55 Cancri e. 

In the case of very long scans, we have to take into account the geometric distortions (dispersion variations across the scanning direction and inclined spectrum) and the positional shifts (horizontal and vertical), as their effect on the structure of the spatially scanned spectrum becomes significant. Especially for fast scans, we found that the vertical shifts are as important as the horizontal ones, because they are coupled with the reading process of the detector (up-stream/down-stream effect), causing exposure time variations. We also found that the time-dependent, long-term systematics, appear to have a different behavior per wavelength channel.

The observed spectrum was analyzed with the \taurex retrieval code. We identified a few important facts: 
\begin{enumerate}
	\item The planet has an atmosphere, as the detected spectral modulations are 6$\sigma$ away from a straight line model.
	\item The atmosphere appears to be light-weighted, suggesting that a significant amount of hydrogen and helium is retained from the protoplanetary disk.
	\item There is no evidence of water vapor.
	\item The spectral features at 1.42 and 1.54 $\mu$m can best be explained by HCN, with a possible additional contribution of other molecules, such as CO, CO$_2$ and C$_2$H$_2$.
	\item This scenario is consistent with a carbon-rich atmosphere (e.g. C/O ratio = 1.1) dominated by carbon bearing species. The model for such an atmosphere was computed independently.
\end{enumerate}

While our results have important implications to the study of 55 Cancri e and other super-Earths, further spectroscopic observations in a broader wavelength range in the infrared are needed to confirm our conclusions.

\begin{acknowledgements}

This work was supported by STFC (ST/K502406/1) and the ERC projects ExoLights (617119) and ExoMol (267219). O.V. acknowledges support from the KU Leuven IDO project IDO/10/2013 and from the FWO Postdoctoral Fellowship programme.

\end{acknowledgements}

{\small
\bibliographystyle{apj}
\bibliography{wfc3_55cnce}
}

\end{document}